\documentclass[letterarticle]{lpor2014}
\usepackage{mathptmx}
\usepackage{amsmath}
\usepackage{cancel}
\usepackage{eqnarray}
\usepackage{soul}
\usepackage{microtype}
\usepackage{newtx}

\begin{document}

\abstract{We present the characterization of ultrashort laser pulses by using the plasma-induced frequency resolved optical switching (PI-FROSt) technique, implemented in ambient air. This recently developed method allows for a temporal reconstruction of a pulse at its focal spot by utilizing a moderately intense pump laser pulse for generating a ionization-induced ultrafast defocusing lens. When propagating through the produced plasma lens, the probe beam to characterize experiences an increase of its size in the far field. The spectrum of the defocused probe field, measured as a function of the pump-probe delay, allows for a comprehensive characterization of the temporal and spectral attributes of the pulse. We report herein the ability of this technique, initially designed for use in rare gases, to operate in ambient air conditions with similar performance. The method is remarkably straightforward to implement and requires no additional optical component other than a focusing mirror, while delivering laser pulse reconstructions of high reliability.}

\title{Temporal characterization of laser pulses using an air-based knife-edge technique}

\author{Pierre B\'ejot\inst{1,*}, Rishabh Kumar Bhalavi\inst{1,2}, Adrien Leblanc\inst{3}, Antoine Dubrouil\inst{2},  Franck Billard\inst{1}, Olivier Faucher\inst{1}, and Edouard Hertz\inst{1}}
\authorrunning{P. B\'ejot et al.}
\mail{pierre.bejot@u-bourgogne.fr}

\institute{$^{1}$ Laboratoire Interdisciplinaire CARNOT de Bourgogne, UMR 6303 CNRS-Universit\'e de Bourgogne, BP 47870, 21078 Dijon, France.\\ $^{2}$ Femto Easy, Batiment Gienah, Cité de la Photonique, 11 avenue de Canteranne, 33600 Pessac France.\\ $^{3}$ Laboratoire d’Optique Appliquée, Ecole Polytechnique, ENSTA, CNRS, Université Paris Saclay, Palaiseau, France.}

\maketitle
\section{Introduction}
After over three decades of continuous development in ultrafast laser technologies, a wealth of diagnostic tools has emerged for the characterization of femtosecond optical pulses \cite{FROG,SPIDER,SRSI,ChirpScan,MIIPS,DESIRE,TIGER}. For an intensive review of this topic, we invite the reader to refer to \cite{Trebino,Walmsley}. In this context, nearly all optical devices designed for pulse characterization require the use of transmissive optics (such as nonlinear crystals, lenses, polarizers, thin glass pieces, and so forth), which can potentially introduce undesired effects on the pulse measurement. For instance, transmissive optics inherently imparts additional spectral phase (which can be nevertheless limited by minimizing the total thickness of the optics) to the pulse under examination, potentially posing challenges, especially for ultra-broadband laser fields measurements. Moreover, in the case of intense laser pulses, transmissive optics may introduce a nonlinear temporal phase due to nonlinear effects or, in the worst scenario, may be subject to optical damage. Lastly, an optical characterization device does not provide the temporal profile of the laser pulse at the exact location where experiments are carried out. Specifically, in pump-probe experiments, the critical pulse characteristics are those at the point where the pump and probe interact, namely, at their focal positions. Recently, a characterization method directly working in air has been developed \cite{TIPTOE}. This technique, called tunneling ionization with a perturbation for the time-domain observation of an electric field (TIPTOE), allows for the direct time sampling of the field to characterize at the focal point. However, since this technique has to resolve the carrier frequency oscillations of the field, it requires to acquire a signal with a sub-cycle resolution. Moreover, the approach can only be applied for moderately chirped input pulses \cite{TIPTOE2}. Recently, we demonstrated that photo-induced free electrons left in the wake of a moderately intense laser pump can be advantageously exploited for characterizing the temporal properties of a pulse \cite{PiFROST_OL}. As recently shown in \cite{LeblancFROST}, the key idea of this phase-matching free method was to produce a temporal analogue of the knife-edge technique widely used for determining the spatial intensity distribution of a beam. When created by a bell-shaped pump beam, a plasma distribution is known to act as a negative lens, simply because the refractive index modification induced by free electrons is negative \cite{CrossDefoc,CrossDefoc2}. As a consequence, when propagating in this low-density plasma, a probe beam will experience a defocusing leading to an increase of its size in the far field. In the time domain, since the plasma is created almost instantaneously by the pump and provided that its lifetime (typically tens to hundreds picoseconds) is longer than the probe duration, only the trailing edge of the probe will be defocused. Combined with a coronagraph placed in the far field so as to obstruct the probe path when it propagates alone, the induced-plasma then acts as a switch that can be viewed as a temporal blade. More particularly, it was shown that measuring the spectrum of the signal propagating around the coronagraph as a function of the pump-probe delay allows for a comprehensive retrieval of the temporal and spectral characteristics of the probe field. This approach, called plasma-induced frequency resolved optical switching (PI-FROSt), features a number of remarkable assets. It is straightforward to implement, free from phase-matching issues, can operate over an exceptionally broad spectral range, in both self- or cross-referenced configurations, at ultra-high repetition rates with no damage threshold \cite{PiFROST_OL}. In order to assess the performance of the method, a noble gas (argon) was used during our first demonstration.
\begin{figure*}[t!]
\begin{center}
\includegraphics[width=17cm,keepaspectratio]{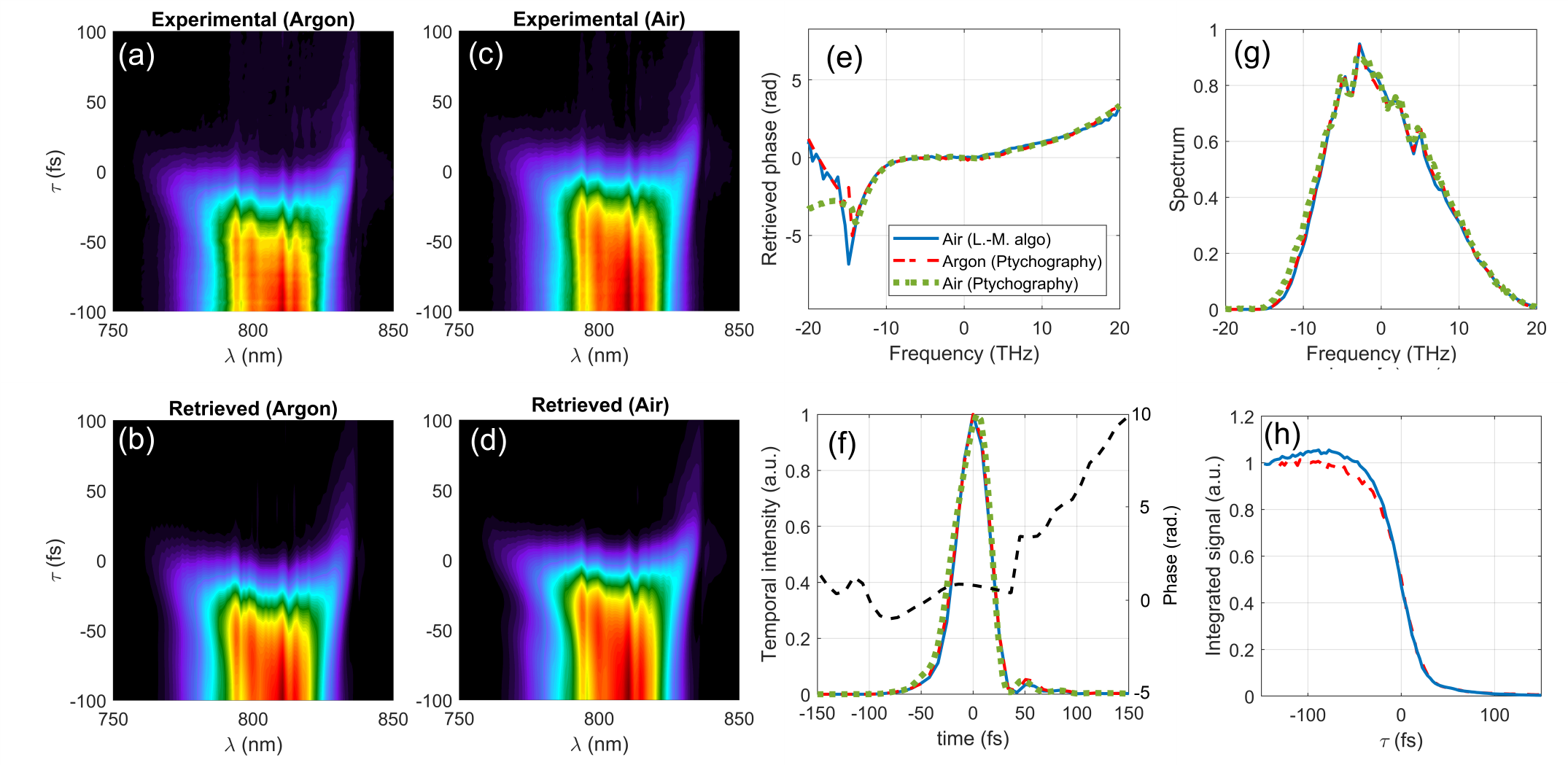}
\caption{Experimental (a) and retrieved (b) PI-FROSt signal obtained in argon compared to the one measured (c) and retrieved (d) in air for a 35\,fs compressed pulse. Spectral phase (e), spectral intensity (g) and temporal intensity profile (f) retrieved by PI-FROSt measurements in the case of argon (red dashed line) and air (green dotted line) using a ptychographic algorithm. The temporal phase retrieved in air from ptychography is depicted as dashed black line. The blue solid line corresponds to retrieval data obtained in air when a standard Levenberg Marquardt algorithm is used (see text). Spectrally integrated signal as a function of the pump probe delay (h) in the case of argon (red dashed line) and air (solid blue line). Both curves are normalized with the same factor for an accurate comparison.}
\label{Fig1}
\end{center}
\end{figure*}
Such a choice simplifies the excitation scheme by avoiding the occurrence of Raman effect. However, while using a static cell allows for controlling the gas composition and pressure, the input window nevertheless introduces additional group velocity dispersion, which can be detrimental for the measurement of few-cycle pulses. In this article, we demonstrate that PI-FROSt measurements can be carried out directly in ambient air. In particular, comparing the results obtained in air with those in argon, it is shown that the Raman-induced molecular alignment \cite{CrossDefoc2} taking place in air does not impact the retrieval process, making the technique extremely convenient to use. 
\section{Cross-defocusing in molecular samples}
A probe pulse, interacting with a rather intense pump in a gas medium, undergoes cross-defocusing that does not solely stem from the plasma. Different other effects, all coming from the nonlinear interaction with the pump, can contribute to the probe size modification in the far-field. For instance, it is known that, far from two-photon resonances, the electronic Kerr effect tends to induce an instantaneous focusing lens. In the present experiment, it means that, when temporally overlapped with the pump beam, the probe beam experiences a cross-focusing effect in addition to the plasma defocusing. If the pump and probe polarizations are parallel (resp. orthogonal), the refractive index change $\Delta n_{\textrm{Kerr}}$ is proportional to $n_2I_\textrm{p}$ (resp. $n_2I_\textrm{p}/3$), where $n_2$ is the nonlinear refractive index of the medium and $I_\textrm{p}$ is the pump intensity. Since the defocusing signal is proportional to the square of the refractive index, the use of orthogonally polarized fields reduces by almost one order of magnitude (a factor 9) the Kerr contribution, which can be neglected in standard PI-FROSt measurements \cite{PiFROST_OL}. In addition to the instantaneous Kerr effect, if the considered molecule does not exhibit a spherical top symmetry (which is the case for oxygen and nitrogen), molecular alignment (i.e., the so-called Raman effect) can take place. For a pulse duration significantly shorter than the rotational period, as with a femtosecond laser pulses, the interaction prepares a rotational wavepacket in the ground state of the molecule by impulsive stimulated Raman transitions. The quantum beatings of the wavepacket manifests itself by a molecular alignment taking place shortly after the pulse turn-on, followed later by periodic and transient revivals of alignment \cite{Seideman,Rosca-Pruna2001}. This molecular alignment also results in a modification of the refractive index experienced by a probe that propagates in the wake of the pump \cite{Renard2003}. Calling $\theta$ the angle between the laser pump polarization direction and the molecular axis, the refractive index change $\Delta n_{\textrm{align}}$ experienced by the probe is proportional to $\langle\langle\textrm{cos}^2\theta\rangle\rangle-1/3$, where $\langle\langle\textrm{cos}^2\theta\rangle\rangle$ denotes the quantum and thermal average of the operator \cite{NicolasAlignement}. Similar to the Kerr effect, the modification of the refractive index induced by molecular alignment depends on the relative polarization of the probe in relation to that of the pump. More particularly, one has $\Delta n_{\textrm{align}_{\perp}}=-\frac{\Delta n_{\textrm{align}_{//}}}{2}$ for a probe polarized perpendicularly with respect to the pump field \cite{Renard2003}. As mentioned, once produced, the generated molecular wavepacket continues to evolve after the pump is turned off, resulting in a periodic, field-free re-alignment of the molecules.
\begin{figure}[htbp!]
\includegraphics[width=8.2cm,keepaspectratio]{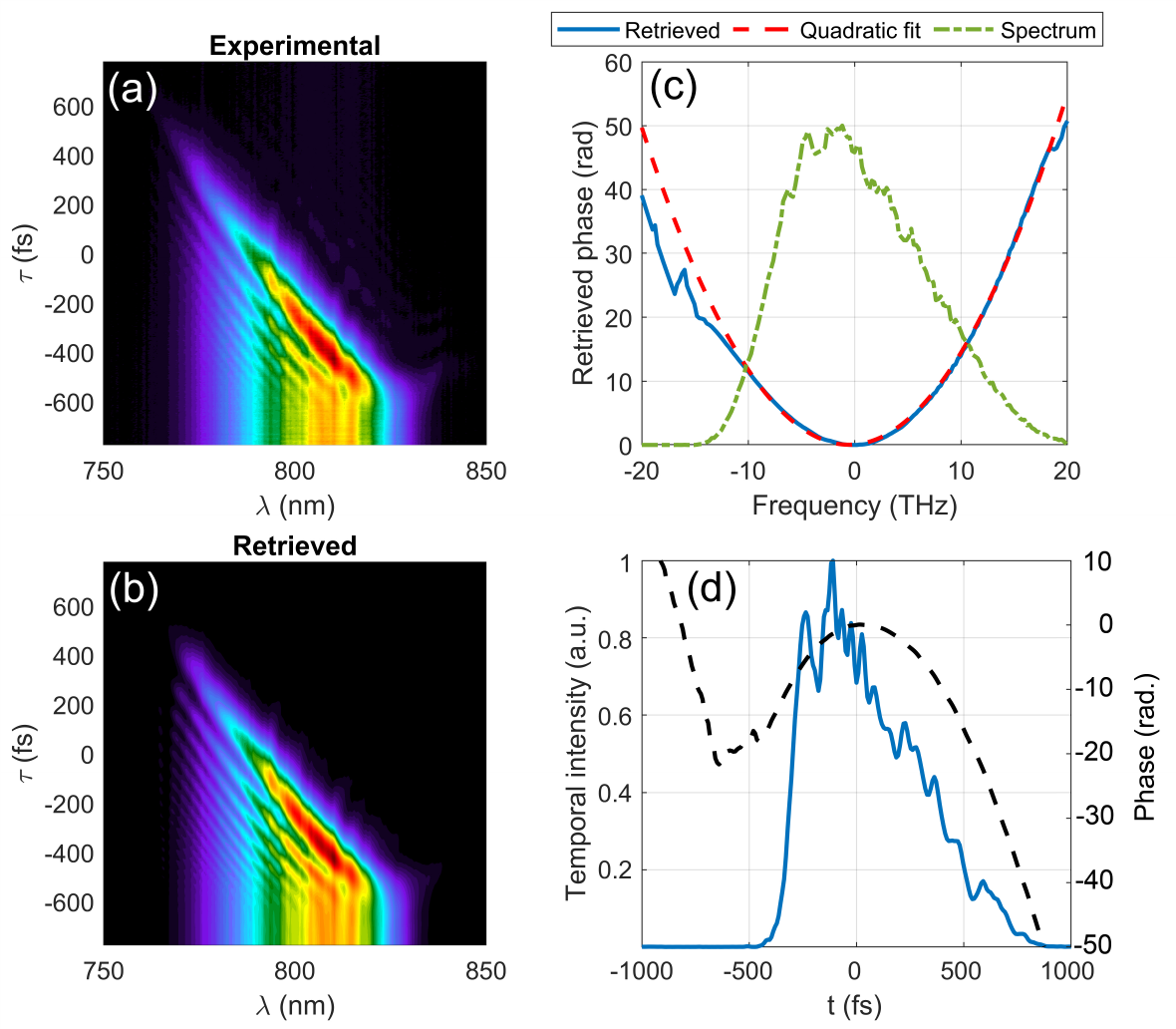}
\caption{Experimental (a) and retrieved (b) PI-FROSt signal obtained in air for a probe field chirped with a 35\,mm SF$_{11}$ flat window. (c) Retrieved spectral intensity (green dot-dashed line), spectral phase (solid blue line) and its associated quadratic fit (dashed red line). (d) Retrieved temporal intensity (solid blue) and phase profiles (dashed black).}
\label{Fig2}
\end{figure}
Throughout the wavepacket evolution, the molecular system evolves from a state where it is preferentially aligned along the pump polarization direction ($\langle\langle \textrm{cos}^2\theta\rangle\rangle>1/3$) to a planar delocalization state where it is primarily contained in the plane perpendicular to the pump polarization ($\langle\langle \textrm{cos}^2\theta\rangle\rangle<1/3$). As a consequence, a probe with the same (respectively, orthogonal) polarization as the pump will undergo either additional focusing (respectively, defocusing) or defocusing (respectively, focusing) when the molecules are aligned or delocalized. For moderate pump intensity, molecular alignment induces a refractive index change proportional to the pump intensity \cite{CrossDefoc}. Note that the rotational Raman contribution to the defocusing signal for an orthogonally polarized probe field (as implemented in the present work) is decreased by a factor four as compared to a parallel configuration. In the context of PI-FROSt measurements, molecular alignment induced by the bell-shaped pump beam therefore results in a supplemental contribution to the cross-defocusing signal \cite{CrossDefoc,CrossDefoc2}. More particularly, since pump and probe are orthogonally polarized, the inertial alignment following the pump pulse leads to a negative contribution to the refractive index experienced by the probe that will add to that of the plasma, thus increasing the defocusing signal just after the pump turns off.  
\section{Experimental results}
In order to evaluate the impact of molecular alignment on PI-FROSt measurements, the signal obtained in air was compared to that recorded in argon under the same configuration. To achieve this, all experiments were conducted with a cell filled with either air or argon (at 1 bar), thus ensuring that the probe pulse experiences exactly the same total group dispersion delay in both cases. The experimental setup is identical to that reported in \cite{PiFROST_OL}. The measurements have been performed on a Ti:Sa femtosecond laser delivering pulses centered at 802\,nm at 1\,kHz and duration 35~fs. The present demonstration was carried out with pump and probe pulses with the same central wavelength and crossed polarizations so as to minimize both Kerr and molecular alignment contributions. The pump (resp. probe) energy was set to 40\,$\mu$J (resp. 10\,nJ). Figure\,\ref{Fig1}(a) depicts the spectrogram $S(\lambda,\tau)$, with $\tau$ the delay of the pump pulse relative to the probe, obtained in argon for a compressed probe pulse. This last exhibits, as already observed in \cite{PiFROST_OL}, an abrupt increase after the pump turns off ($\tau\approx 0$) together with a significant spectral broadening. At this delay, only the falling edge of the probe is diffracted by the plasma, leading to a diffracted pulse with a duration shorter than the initial pulse (manifested as broadening in the spectral domain). Such a broadening provides evidence that the probe pulse has undergone an ultrafast dynamic temporal truncation. The spectrally-integrated signal, displayed as red-dashed line in Fig.\ref{Fig1}(h), monotonically increases when decreasing the delay (from $\tau \approx 0$) in line with the gradual increase of free electron density following the interaction with the pump pulse. This signal does not exhibit evidence of any instantaneous electronic Kerr response, which would manifest as a slight decrease of the signal around $\tau=0$ (since the Kerr contribution has an opposite effect as compared to the plasma), in line with our previous measurements \cite{PiFROST_OL}. Figure\,\ref{Fig1}(c) depicts the spectrogram measured in ambient air for the same experimental conditions. A comparison with Fig.\,\ref{Fig1}(a) reveals that the spectrograms acquired in both gases are highly similar, then suggesting a relatively minor contribution of molecular alignment in the measured signal. The sole qualitative distinction between the two gases is observable when comparing the spectrally-integrated signals [Fig.\,\ref{Fig1}(h)]. While the integrated signal recorded in argon monotonically decreases with increasing delay, that measured in air exhibits an initial increase (in the delay region near $\tau\approx -50$~fs) followed by the same decrease. This difference arises from the (time-delayed) alignment of N$_2$ and O$_2$ molecules. Indeed, as explained above, as the pump and probe are cross-polarized, the alignment introduces a negative refractive index contribution that adds to the plasma contribution, thereby explaining the observed signal increase. 
%
%
In order to reconstruct the probe pulse from these spectrograms, a ptychographic algorithm \cite{LeblancFROST} was first used. The latter enables simultaneous fitting of both the probe spectral characteristics and $\Delta n$ (i.e., the switch) without any assumptions regarding the functional form of the latter. The retrieved spectrograms [see Figs.\,\ref{Fig1}(b,d)] allow to extract, for both gases, the spectral and temporal pulse profiles [depicted as dashed and dotted lines in Figs.\,\ref{Fig1}(e-g)]. The retrieved intensity profile [Fig.\,\ref{Fig1}(f)] reveals a pulse duration close to the expected Fourier-limited 35~fs (full width at half maximum) with a small residual asymmetry as already observed in \cite{PiFROST_OL}. More interestingly, the reconstructions with the ptychography algorithm for air and argon show an excellent agreement, with no discernible differences on the retrieved temporal profiles. Furthermore, we have confirmed that the output of the ptychography algorithm accurately reproduces the integrated spectrum of Fig.\,\ref{Fig1}(h) (not shown for clarity) and particularly the additional molecular contribution observed in air.
\begin{figure}[htbp!]
\includegraphics[width=8.2cm,keepaspectratio]{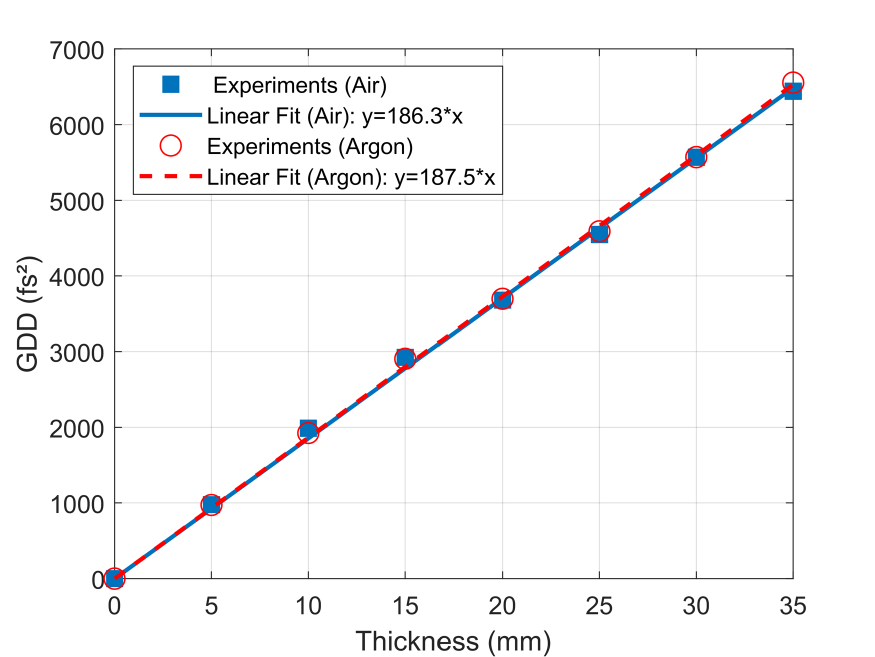}
\caption{Retrieved group delay dispersion as a function of the thickness of the SF$_{11}$ flat window inserted in the probe path for air (blue squares) and argon (red circles) with their associated linear fits.}
\label{Fig3}
\end{figure}
Theses findings indicate that the ptychography algorithm suitably accounts for the residual Raman effect contribution with no impact on the retrieval procedure so that PI-FROSt can be safely conducted in ambient air. Similar to the procedure outlined in \cite{PiFROST_OL}, a reconstruction using a standard Levenberg Marquardt algorithm with a temporal switch of predefined functional shape can also be implemented. For measurements in atomic gases, neglecting the Kerr contribution, the signal writes as:   
\begin{eqnarray}
\begin{aligned}
    &S(\omega,\tau)\propto\left|\int_{-\infty}^{+\infty} E_{\textrm{probe}}(t)\Delta n\left(t-\tau\right)e^{i\omega t}\,dt\right|^2,&\\
    &\Delta n(t)\propto\int_{-\infty}^tI_{\textrm{pump}}^{K}(t')\,dt',&
\end{aligned}
\label{EqAlgo}
\end{eqnarray}
where $\omega$ is the angular frequency, $I_{\textrm{pump}}$ is the pump intensity and $K$ is the effective ionization nonlinearity \cite{PiFROST_OL}. For molecular gases such as air, the switch function $\Delta n(t)$ in Eq.\,\ref{EqAlgo} should also include the molecular contribution. Nevertheless, our analysis led us to the conclusion that the rotational response could be disregarded with no impact on the retrieval procedure. A pulse reconstruction using the Levenberg-Marquardt algorithm and based on Eq.\,\ref{EqAlgo} is displayed in blue solid line of Figs.\,\ref{Fig1}(e-g) for the case of air [Fig.\,\ref{Fig1}(c)]. As evidenced, the retrieved pulse is in excellent agreement with that provided by the ptychography algorithm. This observation demonstrates that, despite its residual presence in the integrated signal, the Raman effect has no impact on the retrieval procedure. We point out that similar conclusions have been obtained for various input chirps or pump energies. At first glance, one might assume that omitting the Raman contribution in Eqs.\,\ref{EqAlgo} could lead to a significant error in the retrieval procedure. As a matter of fact, the observed robustness can primarily be attributed to the fact that the contribution of molecular alignment is time-delayed and thus does not alter the region of temporal cutting, which plays the dominant role in pulse reconstruction. The redundancies in the information gathered from the FROSt spectrogram make the spectral phase reconstruction procedure extremely robust.
\begin{figure}[htbp!]
\includegraphics[width=8.2cm,keepaspectratio]{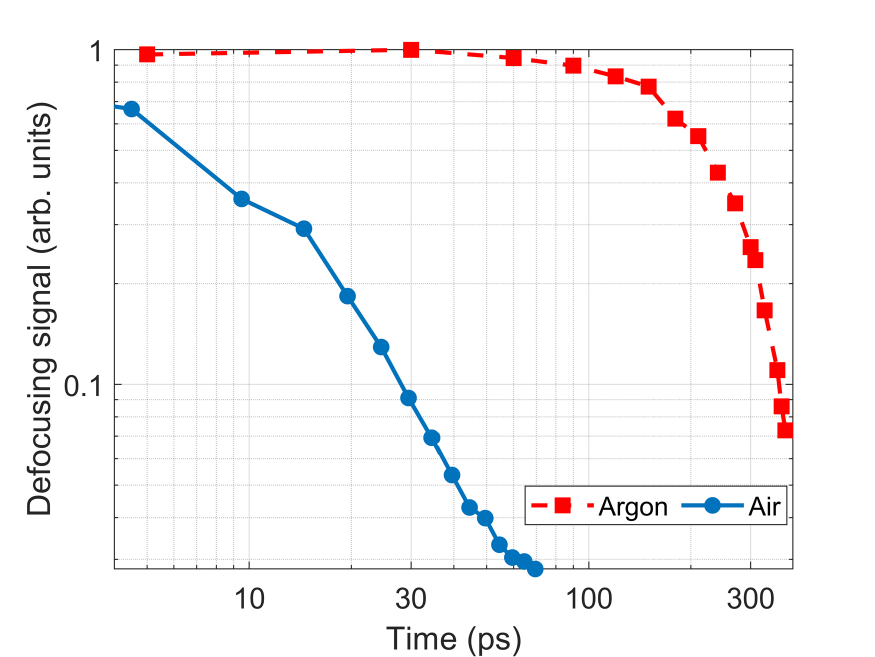}
\caption{Spectrally-integrated PI-FROSt signal as a function of pump-probe delay obtained in air (blue circles) and in argon (red squares).}
\label{Fig4}
\end{figure}
In other words, the algorithm refrains from modifying the reconstructed probe spectrum to better adjust for the Raman contribution of the measured signal because doing so would lead to significant mismatches in other regions of the spectrogram.
In order to evaluate the accuracy of the PI-FROSt technique, we inserted several flat SF$_{11}$ windows with varying thicknesses $d$ (ranging from 5\,mm to 35\,mm) along the probe path. For each window, PI-FROSt measurements were performed both in air and argon to assess the spectral phase of the probe. A typical measurement in ambient air is depicted in Fig.\,\ref{Fig2} for $d=35$~mm. As already observed \cite{PiFROST_OL}, the chirp induced by the plate leads to a pronounced spectral asymmetry in the cutoff region, the blue components being diffracted before the red ones, as a result of the linear frequency chirp. In order to evaluate the additional phase introduced by the windows, we subtracted the phase $\phi(\omega,d=0)$ previously obtained without any window [see Fig.\,\ref{Fig1}(e)] from the retrieved phase $\phi(\omega,d)$. The resulting phase difference was then fitted with a second-order polynomial in $\omega$, enabling estimation of the group dispersion delay caused by the windows. The results of this procedure is summarized in Fig.\,\ref{Fig3}. As shown, the retrieved group-delay dispersion linearly increases with $d$, the results obtained in air (blue circles) and argon (red squares) being highly consistent. A linear fit of the obtained curves allows to evaluate the group velocity dispersion of SF$_{11}$. The latter is found to be 186.3\,fs$^2$/mm (resp. 187.5\,fs$^2$/mm) for the measurements performed in air (resp. argon), which is very close to the expectation (186.7\,fs$^2$/mm). We stress that these findings are particularly noteworthy given that the signal measurements in both air and argon, under identical dispersion conditions, were not conducted sequentially (the measurements were initially performed in air across various thicknesses of dispersive plates before being replicated in argon). The remarkable similarity observed in the pulse reconstructions for the two gases [see for instance Figs.\,\ref{Fig1}(e-f)] constitutes a further evidence of the robustness of the PI-FROSt method.
Finally, the PI-FROSt signal lifetime was assessed by measuring the integrated signal as a function of the pump-probe delay across a wide temporal range. Here again, the measurements were performed both in argon and air. The evaluation of the ionization relaxation is of prime importance for a potential applicability of the method for ultra-high repetition rates (for instance, for MHz repetition rate high-power Ytterbium lasers). As shown in Fig.\,\ref{Fig4}, the FROSt signal observed in air vanishes in less than 100\,ps, while in argon, it demonstrates significantly longer persistence, exceeding 350\,ps, which was the upper limit achievable with our translation stage. This rapid decay of the PI-FROSt signal in the case of air opens up the possibility of conducting measurements at very high repetition rates, potentially reaching several gigahertz.
\section{Conclusion}
In this paper, we have demonstrated the applicability of the plasma-induced frequency-resolved optical switching technique in ambient air, extending its prior application to noble gases. Through a comprehensive comparison of results obtained in both argon and air, we have established that the contribution of the rotational alignment to the total refractive index  experienced by the probe does not compromise the fidelity of the retrieval process when using a ptychography algorithm but also when using a standard Levenberg Marquardt algorithm with a temporal switch of predefined shape (disregarding the molecular Raman response). The capability of PI-FROSt to operate in ambient air alleviates the necessity for a gas cell, rendering the technique extremely straightforward to implement. Furthermore, since no transmissive optic is necessary prior to the pump-probe interaction, the method does not influence the spectral phase of the pulse being measured and is immune to the issue of phase matching constraints.
\begin{acknowledgement}
The authors acknowledge financial support of French programs "Investments for the Future" operated by the National Research Agency (ISITE-BFC, contract ANR-15-IDEX-03; EIPHI Graduate School, contract ANR-17-EURE-0002; EQUIPEX+ Smartlight, contract ANR-21-ESRE-0040), from Bourgogne Franche-Comt\'e region and European Regional Development Fund.
\end{acknowledgement}

\end{document}